\documentclass[aps,prd,tightenlines,superscriptaddress,nofootinbib]{revtex4}
\usepackage{epsfig}
\usepackage{graphicx}
\usepackage{psfrag}
\usepackage{amsmath,amssymb}
\usepackage{colordvi}
\usepackage{amsfonts}
\usepackage{enumerate}
\usepackage{slashed}
\usepackage{color}
\usepackage{xcolor}

\begin{document}
\title{Comment on ``Next-to-leading order forward hadron production in the small-x regime: rapidity factorization" arXiv:1403.5221 by Kang {\it et al.}}
\author{Bo-Wen Xiao}
\affiliation{Key Laboratory of Quark and Lepton Physics (MOE) and Institute
of Particle Physics, Central China Normal University, Wuhan 430079, People's Republic of China}
\author{Feng Yuan}
\affiliation{Nuclear Science Division, Lawrence Berkeley
National Laboratory, Berkeley, California 94720, USA}
\date{\today}
\vspace{0.5in}
\begin{abstract}
In a recent paper~\cite{Kang:2014lha}, Kang {\it et al.} proposed the so-called 
``rapidity factorization'' for the single inclusive forward hadron production in $pA$ collisions. 
We point out that the leading small-$x$ logarithm was mis-identified in this paper, 
and hence the newly added next-to-leading order correction term 
is unjustified and should be absent in view of the small-$x$ factorization. 
\end{abstract}

\maketitle

Single inclusive hadron production in the forward $pA$ collisions is one of the simplest and most interesting
processes which can probe the onset of gluon saturation in dense nuclear targets. 
The leading order formula for this process was derived in Ref.~\cite{Dumitru:2002qt} in 2002.
The next-to-leading order (NLO) corrections were calculated in the small-$x$ factorization formalism 
in Refs.~\cite{Chirilli:2011km,Stasto:2013cha} a few years ago. 
 In a recent publication~\cite{Kang:2014lha}, the NLO result
was re-derived following the same method in Ref.~\cite{Chirilli:2011km}, and 
reproduced what have been computed 
(including the collinear factorization for the parton
distribution and fragmentation functions), except for
the small-$x$ factorization part. In lieu of the small-$x$ factorization scheme used in Ref.~\cite{Chirilli:2011km},
Ref.~\cite{Kang:2014lha} proposed the so-called ``rapidity factorization''.
We disagree with the rapidity factorization argument used in Ref.~\cite{Kang:2014lha}
which leads to a new NLO correction. 
We believe that this comes from a mis-identification of leading
logarithms at small-$x$ in their calculations. 

First of all, as a general remark, in the final result of Ref.~\cite{Kang:2014lha},
the hadronic kinematic variable $Y=\ln\left(\frac{s}{m_p^2}\right)$ appears. Here, 
$s$ and $m_p$ are the center of mass energy squared and the proton mass. 
This violates the generic factorization principle in parton physics, where the hadronic cross
section should be written as a convolution of parton distribution and
the partonic cross section. The latter does not depend on the hadrons' momenta. 
If this does not hold, it means that there is no factorization at all. As we have 
already demonstrated 
the factorization for this process in Ref.~\cite{Chirilli:2011km}, 
the hadronic variable $Y$ never enters in the factorization formula for single inclusive hadron production. It appears
in the result of Ref.~\cite{Kang:2014lha} due to a mis-identification of the
leading logarithms of small-$x$ resummation. We will elaborate
more on this in the following.
In addition, the appearance of the proton mass in the result of \cite{Kang:2014lha}
cast strong suspicion from the perturbative calculation point of view, since
proton mass has normally be regarded as nonperturbative scale. 
How proton mass enters in a perturbative calculations needs further justification.

The new NLO correction in the so-called ``rapidity factorization'' arises from the 
mis-identification of the large logarithms associated
with small-$x$ evolution in Ref.\cite{Kang:2014lha}. In the following, we 
show the correct evaluation of the large logarithms associated with the small-$x$
physics, and demonstrate how to obtain the consistent result in the spirit of factorization. 
These derivations have been clearly shown in Ref.~\cite{Chirilli:2011km, Mueller:2012uf}.
To emphasize the conceptual difference, we elaborate these arguments
step by step as follows.

We take the quark channel contribution as an example. The leading
contribution can be formulated as the quark scattering on nucleus target and
fragmenting into a final state hadron, and the cross section is written
as 
\begin{equation}
\frac{d\sigma(pA\to h+X)}{dyd^2p_\perp}=\int q(x_p)\otimes D_q(z)\otimes {\cal F}_{x_g}(r_\perp) \ , \label{qq}
\end{equation}
where $x_p=k_\perp e^y/\sqrt{s}$, $k_\perp=p_\perp/z$, 
and $x_g=k_\perp e^{-y}/\sqrt{s}$ representing the $x$-value at which the dipole amplitude is
evaluated for the quark production. 

Although it was not explicitly written, from their calculation, 
the factorization that Ref.~\cite{Kang:2014lha} claimed seems take the 
following form,
\begin{equation}
\frac{d\sigma(pA\to h+X)}{dyd^2p_\perp}|_{Ref.[1]}=\int q(x_p,\mu)\otimes D_q(z,\mu)\otimes {\cal F}_{Y_0}(r_\perp)
\left[1+\alpha_s\int_{Y_0}^{Y}dY' {\rm BK}\otimes +\alpha_s({\rm other~terms}) \right] \ , \label{kang1}
\end{equation}
for incoming quark channel contribution, where $y$ and $p_\perp$
are rapidity and transverse momentum of final state hadron, $q(x_p,\mu)$ and $D_q(z,\mu)$
are the integrated quark distribution from the proton
and fragmentation function for the final state hadron, respectively. ${\cal F}_{Y_0}(r_\perp)$ 
is defined as the dipole amplitude from nucleus, and $\alpha_s({\rm other~terms})$ 
represent those contributions which are not related to small-$x$ evolution. 
Since we focus on the small-$x$ factorization here, we omit
those terms and simplify the formula for convenience. To obtain the differential cross section
depending on transverse momentum, the Fourier transformation is performed. 
The above large logarithmic correction in terms of $\alpha_s (Y-Y_0)$ ($Y-Y_0=\ln\left(sx_g/m_p^2\right)$
with $Y_0=\ln(1/x_g)$ and $x_g=p_\perp e^{-y}/z\sqrt{s}$)
can be absorbed into the dipole amplitude ${\cal F}(r_\perp)$.
Therefore, at the level of the leading logarithmic approximation, their result
can be written as
\begin{equation}
\frac{d\sigma(pA\to h+X)}{dyd^2p_\perp}|_{Ref.[1]}=\int q(x_p,\mu)\otimes D_q(z)\otimes {\cal F}_{Y=\ln(s/m_p^2)}(r_\perp)
\left[1+\alpha_s({\rm other~terms}) \right] \ ,
\end{equation}
where the $\alpha_s$ correction does not contain any large logarithms associated
with small-$x$ evolution. This clearly contradicts with the concept of factorization as well as Eq.~(\ref{qq}),
where leading order result depends on $x_g$, instead of $\ln(s/m_p^2)$~\footnote{In any application of
factorization in hadronic process, the leading order perturbative
calculation represents the leading logarithmic resummation, under the assumption
that the factorization is valid and the relevant evolution equation is known. 
This applies to the case discussed here.}. The above result also means that the dipole amplitude 
is independent of the rapidity of produced hadrons $y$. It would be a universal function only depending on total energy $s$,
meaning that one only needs one single dipole amplitude for RHIC and LHC 
separately. The direct consequence is that the gluon saturation scale does not depend on the 
rapidity of the produced hadron either. On the other hand, the dipole amplitude ${\cal F}(r_\perp)$, 
which is derived from the scattering amplitude between the quark and the target nucleus, 
should not depend on the total centre of mass energy $s$.

In our opinion, the gluon rapidity $y_g\equiv\ln\frac{1}{1-\xi}$ \footnote{$(1-\xi)$ is the 
longitudinal momentum fraction that the radiated gluon carries with respect to the 
parent quark.} in Ref.~\cite{Kang:2014lha} by definition should be the rapidity 
separation between the radiated gluon and the parent \textit{quark}, instead of 
the projectile \textit{proton}. Only in very forward production, are these two quantity 
the same. But conceptually they differ by a factor of $\ln\frac{1}{x_p}$. Thus, if one 
defines $y_A=Y-y_g$ as the rapidity of the radiated gluon w.r.t. the target nucleus, 
$Y$ should be the rapidity interval between the quark and target nucleus, which is, 
in principle, the same as $Y_0\equiv\ln\frac{1}{x_g}$.
Therefore, the new finite correction introduced in Ref.~\cite{Kang:2014lha} should 
be identically zero.

As we have shown in Ref.~\cite{Chirilli:2011km},
the choice of rapidity interval should reflect the correct leading logarithmic contribution from
gluon radiation at small-$x$. Here, we elaborate in more details, and
concentrate on large logarithms from small-$x$ evolution from 
one gluon radiation. The kinematics is specified as follows: incoming quark with
momentum $p=(p^+,0^-,0_\perp)$ scattering on the nucleus
with momentum $P_A=(0^+,P_A^-,0_\perp)$;  final state quark 
$k=(\xi p^+,k^-,k_\perp)$ and gluon $k_1=((1-\xi)p^+,k_{1}^-,k_{1\perp})$.
The rapidity divergence associated with small-$x$ physics
comes from the kinematic region of the radiated gluon parallel
to the nucleus, leading to the following integral,
\begin{equation}
\frac{\alpha_sN_c}{2\pi}\int_0^1 \frac{d\xi}{(1-\xi)} \ .
\end{equation}
This rapidity divergence appears in both real and virtual graphs. 
To regulate this divergence, a cut-off scheme can be used,
which has to apply to both real and virtual contributions
consistently. In particular, the cut-off in the above integral 
reflects the small-$x$ logarithms, which can be explicitly identified from the kinematics.
According to the on-shell kinematical requirement for the 
radiated gluon: $k_1^2=0$,  one gets $k_1^-=\frac{k_{1\perp}^2}{2(1-\xi)p^+}$.
Due to energy momentum conservation, we have strong
constraints on $k_1^-<P_A^-$. Therefore, $\xi$-integral
in the real diagram is constrained as
\begin{equation}
(1-\xi)>\frac{k_{1\perp}^2}{k_\perp^2}x_g(1+\mathcal{O}(x_g))\ ,
\end{equation}
where $\frac{k_{1\perp}^2}{2p^+P_A^-}=\frac{k_{1\perp}^2}{x_p s}=\frac{k_{1\perp}^2}{k_\perp^2}x_g$ 
is used to arrive at the above expression. Therefore, the rapidity
divergent integral leads to the following large logarithm,
\begin{equation}
\frac{\alpha_sN_c}{2\pi}\ln\left(\frac{1}{x_g}\right) \ ,
\end{equation}
plus terms which are subleading in the small-$x$ resummation, such 
as $\ln({k_\perp^2}/{k_{1\perp}^2}) $.
This is how the large logarithms emerge in the gluon radiation. 
Physically, the small-$x$ evolution resum large logarithms coming 
from collinear gluon radiation with large rapidity difference while
the transverse momentum is the same order. 
Therefore, the leading logarithms in this process is coming from
$\ln (1/x_g)$. It is not in terms of $\ln (s/m_p^2)$. This is essentially the reason that 
the target gluon distribution function is function of $x_g$ instead of function of $\ln (s/m_p^2)$.
Similar analysis has also been applied to obtain
the Sudakov double logarithms in hard processes in $pA$ collisions~\cite{Mueller:2012uf},
where the exact kinematics of the radiated gluon is key to derive
the consistent resummation results.
Of course, a complete factorization should allow us to have freedoms to choose the
rapidity for the dipole amplitude, supplemented with the cancellation
of this rapidity dependence in the final result. 
Technically, this can be
done by choosing another cut-off (such as $(1-\xi)>\delta=e^{-Y_\mu}$) for both the dipole amplitude 
and the cross section calculation. After performing the subtraction, the remaining hard part will
depend on the difference between $Y_\mu$ and $Y_{phys}=\ln(1/x_g)$,
whereas the dipole amplitude depending on $Y_\mu$. In the final factorization
formula, the $Y_\mu$ dependence cancels out. Therefore, $Y_\mu$ can be served
as factorization scale for the rapidity factorization. We can do the same 
computation for any other scheme, such as tilting the Wilson line, and we should be able to obtain the
same results.
Therefore, the small-$x$
factorization leads to the following expression for this part~\cite{Chirilli:2011km},
\begin{equation}
\frac{d\sigma(pA\to h+X)}{dyd^2p_\perp}|_{Ref.[3,4]}=\int q(x_p,\mu) \otimes D_q(z,\mu)\otimes
{\cal F}_{Y_\mu}(r_\perp)
\left[1+\alpha_s\int_{Y_{\mu}}^{Y_{phys}}dY' {\rm BK}\otimes +\alpha_s({\rm other~terms}) \right] \ ,
\end{equation}
where $Y_\mu$ represents the scale separation for the small-$x$
factorization. The above equation is physically different
from Eq.~(\ref{kang1}) which was implied in Ref.~\cite{Kang:2014lha}.

In conclusion, we have pointed out that the leading logarithms 
identified in Ref.~\cite{Kang:2014lha} is not correct and the claimed
rapidity factorization is not consistent with the small-$x$ factorization 
proposed in Ref.~\cite{Chirilli:2011km}.

As a final note. we would like to emphasize that the factorization formula 
we proved in Ref.~\cite{Chirilli:2011km} is only valid in the small-$x$
domain, where the saturation scale $Q_s$ set the hard momentum scale.
This factorization formula will break down at large transverse momentum $p_\perp >Q_s$.
This is exactly what has been shown in the detailed 
numeric studies in Ref.~\cite{Stasto:2013cha}, where it was found that 
the NLO corrections become negative in the transverse momentum
region beyond the saturation scale $Q_s$ for a given rapidity. 
This is because, in this region, the hard gluon radiation becomes dominant contribution,
and thus it is appropriate to apply the collinear factorization to 
calculate the differential cross section, instead of using
the small-$x$ factorization formalism.
Therefore, we shall expect a matching between low and high
transverse momentum region for inclusive hadron production
in $pA$ collisions: in large transverse momentum ($>Q_s$) region,
we apply collinear factorization; while in low transverse momentum ($<Q_s$) region,
we apply small-$x$ factorization, and in between, we should match
between these two calculations. Following this idea, it is found that the transverse
momentum spectrum data from low to high transverse momentum region in $dAu$ collisions at RHIC
can be desribed~\cite{Stasto:2014sea}.

We thank G. Chirilli, Y. Kovchegov, A. Stasto, and D.~Zaslavsky for discussions and comments. 
This work was supported in part by the U.S. Department of Energy under 
the contracts DE-AC02-05CH11231.


\begin{thebibliography}{99}

\bibitem{Kang:2014lha} 
  Z.~-B.~Kang, I.~Vitev and H.~Xing,
  arXiv:1403.5221 [hep-ph].


\bibitem{Dumitru:2002qt} 
  A.~Dumitru and J.~Jalilian-Marian,
 Phys.\ Rev.\ Lett.\  {\bf 89}, 022301 (2002).
  
  \bibitem{Chirilli:2011km} 
  G.~A.~Chirilli, B.~-W.~Xiao and F.~Yuan,
  Phys.\ Rev.\ Lett.\  {\bf 108}, 122301 (2012);
  Phys.\ Rev.\ D {\bf 86}, 054005 (2012).
  
\bibitem{Stasto:2013cha} 
  A.~M.~Stasto, B.~-W.~Xiao and D.~Zaslavsky,
  Phys.\ Rev.\ Lett.\  {\bf 112}, 012302 (2014).
 

   
  
\bibitem{Mueller:2012uf} 
  A.~H.~Mueller, B.~-W.~Xiao and F.~Yuan,
  Phys.\ Rev.\ Lett.\  {\bf 110}, 082301 (2013);
  Phys.\ Rev.\ D {\bf 88}, 114010 (2013).

%
\bibitem{Stasto:2014sea} 
  A.~M.~Staśto, B.~-W.~Xiao, F.~Yuan and D.~Zaslavsky,
  arXiv:1405.6311 [hep-ph].\end{thebibliography}
\end{document}